\documentclass[12pt,notitlepage,a4paper]{article}
\pdfoutput=1
\usepackage[a4paper,margin=1in]{geometry}
\usepackage{amsmath,amssymb,amscd,amsfonts}
\usepackage{mathtools}
\numberwithin{equation}{section}
\usepackage{url}
\usepackage[shortlabels]{enumitem}
\usepackage{tikz} 
\usetikzlibrary{decorations}
\usetikzlibrary{arrows,decorations.markings}         
\usepackage[hidelinks]{hyperref}
\usepackage[open,
  openlevel=2,
  atend,
  numbered]{bookmark}
\usepackage{graphicx}
\usepackage[final]{pdfpages}
\usepackage{threeparttable}
\usepackage[small,sc]{caption}
\usepackage{float}
\usepackage{stackengine}
\usepackage{scalerel}
\usepackage[bf,small,center]{titlesec}
\usepackage{tocloft}

\newlength{\mylen}	
\urldef\footurlb\url{cocalc.com/dfriedan/DM/SM}
\urldef\footurla\url{physics.rutgers.edu/~friedan}
\def\eq{\begin{equation}}
\def\en{\end{equation}}
\def\eqg{\eq\begin{gathered}}
\def\eng{\end{gathered}\en}
\def\eqa{\eq\begin{aligned}}
\def\ena{\end{aligned}\en}

\def\SO{\mathrm{SO}}
\def\SU{\mathrm{SU}}

\def\Spin{\mathrm{Spin}}
\DeclareMathOperator{\cn}{cn}

\def\expval#1{\langle \, #1 \,\rangle}

\def\gauge{\mathrm{gauge}}

\def\tr{\mathrm{tr}}

\def\EW{\scriptscriptstyle\mathrm{EW}}
\def\vH{v}
\def\kB{k_{\scriptscriptstyle \mathrm{B}}}

\def\Higgs{\scriptscriptstyle \mathrm{Higgs}}

\def\CGF{\scriptscriptstyle \mathrm{CGF}}

\def\curvature{\text{curvature}}
\def\GeV{\text{\footnotesize GeV}}
\def\sunit{\text{\footnotesize s}}
\def\epsilonb{\epsilon}
\def\EhatCGF{\hat E_{\CGF}}
\def\ECGF{E_{\CGF}}
\def\ntimes{\,{\times}\,}

\def\gtwo{g}
\def\lambdaH{\lambda}

\stackMath
\def\dyhat{-0.2ex}
\newcommand\myhat[1]{\ThisStyle{%
              \stackon[\dyhat]{\SavedStyle#1}
                              {\SavedStyle\hat{\phantom{#1}}}}}
\def\that{\kern0.1em\myhat{\kern-0.1em t}}                              

\def\eff{{\scriptscriptstyle \mathrm{eff}}}
\def\ssm{{\scriptstyle \mathrm{m}}}
\def\ssc{{\scriptstyle \mathrm{c}}}
\def\ssLambda{{\scriptstyle \Lambda}}

\def\kg{\mathrm{kg}}
\def\munit{\mathrm{m}}
\def\cmunit{\text{cm}}
\def\Msun{\textup{M}_{\odot}}
\def\texttilde{\raise-0.7ex\hbox{\!\texttt{\char`\~}}}
\def\rhob{\rho_{b}}
\def\rb{r_{b}}
\def\mb{m_{b}}

\def\SM{SM}
\def\GR{GR}

\titlespacing{\section}{3pc}{1.75pc}{0.8pc}
\titleformat{\section}
  {\centering\normalsize\bfseries} 
  {\large\thesection}
  {0.5em} 
  {\large}
\begin{document}
\def\title{First principles cosmology 
of
the Standard Model epoch}
%


\begin{center}
{\Large \title}
\vskip5ex
{\large Daniel Friedan}
\vskip2ex
{\it
New High Energy Theory Center
and Department of Physics and Astronomy,\\
Rutgers, The State University of New Jersey,\\
Piscataway, New Jersey 08854-8019 U.S.A. and
\vskip0.5ex
Science Institute, The University of Iceland,
Reykjavik, Iceland
\vskip0.5ex
\href{mailto:dfriedan@gmail.com}{dfriedan@gmail.com}
\qquad
\href{https://physics.rutgers.edu/\textasciitilde friedan/}
{physics.rutgers.edu/\texttilde friedan}
}
\vskip2ex
\today
\end{center}
%
%
\begin{center}
\vskip3ex
{\sc Abstract}
\vskip3ex
\parbox{0.96\linewidth}{
\hspace*{1.5em}
This is a summary of a project
to construct a first principles 
cosmology of the Standard Model epoch, the period starting
shortly before the electro-weak transition.
The cosmology is derived from a simple initial state
entirely within the \SM\ and General Relativity.
The initial state is 
semi-classical ---
concentrated near a 
classical solution of the \SM\ equations of motion
---
and is precisely specified
by a few simple conditions.
The dark matter is a classical effect,
a coherent state of the $\SU(2)$-weak 
gauge field and the Higgs field.
The leading order, classical universe contains only the dark 
matter.  Ordinary matter is a correction due to
the fluctuations of the \SM\ fields
around the classical trajectory.
The initial state produces a homogeneous, isotropic, flat
universe.
There are no adjustable parameters.
No physics beyond the \SM\ is invoked.
Only the classical calculations have been done
so far.
The time evolution of the fluctuations
remains to be calculated.
}
\end{center}

\vspace*{0.5ex}

%
%
%
%
%

\section{Introduction}

The Standard Model cosmological epoch 
is the period starting
at energy scale roughly 1\,TeV
somewhat before the electro-weak transition
The project is to formulate
a simple initial quantum state
whose time evolution
accurately explains the \SM\  epoch
entirely within the \SM\ and classical \GR.
It is a top-down approach to cosmology.  The initial
state is completely
specified by a few simple conditions:
\begin{enumerate}[leftmargin=2cm,rightmargin=2cm,itemsep=0.0ex]
\item The universe is governed by the \SM\ 
and classical \GR\ (with $\Lambda$).
\item The universe is a 3-sphere.
\item The initial state of the \SM\ epoch is semi-classical.
\item The initial state has a certain $\Spin(4)$ symmetry.
\item The initial energy is $ >  10^{107}$ in natural units.
\end{enumerate}
The cosmology is a systematic expansion around a classical solution
of the \SM,
a coherent state of the  Higgs field
and the $\SU(2)$ gauge field
(the cosmological gauge field or CGF).
The CGF acts as a perfect fluid
with $w_{\CGF}\approx 0$.
It is dark matter.
The classical CGF universe contains only the dark matter.
Ordinary matter appears as a sub-leading correction due to fluctuations of the 
\SM\ fields around the classical trajectory.

The classical solution  has the basic 
structure of the \SM\ epoch:
the electro-weak transition followed by an expanding universe that
contains only dark matter and is homogeneous, isotropic, and flat.
This classical CGF universe is the dark matter skeleton of the \SM\ epoch,
to be fleshed out by the fluctuations.
The time evolution of the fluctuations is a perturbative calculation within the \SM\ 
and \GR\ that remains to be done.

Laboratory high energy physics looks for new physics in small discrepancies from 
the \SM.
The idea is to put cosmology in the same situation.
If this project works out,
if the initial condition accounts accurately for the \SM\ epoch,
then any discrepancies will be signs of new physics.
Of course the project may be too ambitious.
It could be that the \SM\ epoch depends on undiscovered
particles and fields.
Still, until those particles or fields are actually discovered,
the top-down approach seems worth trying.

Dark matter being a \SM\ effect
explains why no dark matter particles have been found.
It  predicts that no such particles will be found.
Dark matter being a classical effect and ordinary matter
a sub-leading correction
explains why most 
of the matter in the universe is dark matter.

It seems nontrivial that any simple, natural initial state
should give the basic structure of the \SM\ epoch as first 
approximation,
with a systematic method to calculate corrections.
It might be worthwhile to
calculate the corrections
to see if the details come out right.

The initial condition was proposed in \cite{Friedan:2020poe} .
The classical time evolution was calculated in \cite{Friedan2022:Atheory}
and the dark matter identified.
In \cite{Friedan2022:DMStars} a first step is taken towards 
identifying detectable signals of the dark matter.
In \cite{Friedan2022:Stability} a first step is taken towards constructing the 
initial state of the fluctuations.

\section{The initial state}
The leading order approximation
is a $\Spin(4)$-symmetric classical solution of the \SM\ equations of 
motion.
$\SO(4)$ is the symmetry group of the 3-sphere in euclidean 4-space.
$\Spin(4)=\SU(2){\,\times\,}\SU(2)$ is the simply connected covering 
group.  $\Spin(4)$ acts on spinors on the 3-sphere.
The $\SU(2)$ gauge bundle of the \SM\ is identified with the spinor 
bundle. The U(1) and SU(3) gauge bundles are trivial product bundles
over the 3-sphere.
This defines the action of
$\Spin(4)$ on the space-time metric and 
on the \SM\ fields.

The metric (in $c=1$ units) is
$
ds^{2} = R(\that)^{2}\left(
-d\that{\;\!}^{2} +  \hat g_{ij}(\hat x)d\hat x^{i}d\hat x^{j}
\right)
$
where $\that$ is conformal cosmological time,
$\hat g_{ij}(\hat x)$ is the metric of the unit 3-sphere in 4-space,
and $R(\that)$ is the radius of the universe.
Co-moving time $t$ is given by
$dt = R(\that) d \that$.

The \SM\ fields with nontrivial  classical values
are the Higgs field $\phi$
and the SU(2) gauge field.
The Higgs field is fixed at $\phi=0$ by the symmetry.
The SU(2)  covariant derivative is
$\hat D_{i} = \hat \nabla_{i} + \hat b(\that) \hat \gamma_{i}$
where $\hat \gamma_{i}(\hat x)$ are the Dirac matrices on the unit 3-sphere
and $\hat \nabla_{i}$ is the metric covariant derivative on spinors.
The degree of freedom $\hat b(\that)$
is the cosmological gauge field (CGF).
The Yang-Mills action is
\eq
\label{eq:bhataction}
\frac1\hbar S_{\gauge} =
\frac{6\pi^{2}  }{g^{2}}
\int
\bigg[-\frac12 \bigg(\frac{d\hat b}{d\that}\bigg)^{2}+ \frac12 
(\hat b^{2}-1)^{2}\bigg]
d\that
\qquad
g^{2}=0.426
\en
$g$ is the $\SU(2)$ gauge coupling constant of the \SM\ \cite{PDG2021}.
The conserved quantity
\eq
\EhatCGF = \frac12 \left(\frac{d\hat b}{d\that}\right)^{2} + \frac12(\hat b^{2}-1)^{2}
\en
is the dimensionless energy (in  energy  units $\hbar/R$).
The CGF is an anharmonic oscillator.
The classical equation of motion is solved by a Jacobi
elliptic function
\eqg
\label{eq:classicalsoln}
\hat b(\that) = \frac{ k\cn( z,k)}{\epsilonb }
\qquad
z = \frac{\that}{\epsilonb } 
\qquad
\EhatCGF = \frac1{8\epsilonb ^{4}}
\qquad
k^{2}= \frac12 + \epsilonb ^{2}
\eng
The energy condition
$\EhatCGF > 10^{107}$, $\epsilonb < 10^{-27}$
will come later as a physical condition
needed to produce the observed flatness of the present universe.

Scale the coordinates and degrees of freedom by $\epsilon$,
\eqg
x^{0} = z = \frac{\that}{\epsilon}
\qquad
x^{i} =\frac{\hat x^{i}}{\epsilon}
\qquad
a(z) = \epsilon R(\that)
\qquad
b(z) = \epsilon \hat b(\that) =  k\cn( z,k)
\eng
The metric and Dirac matrices become
\eqg
ds^{2} = a(z)^{2}\left(
-dz^{2} +   g_{ij}( x)d x^{i}d x^{j}
\right)
\qquad
\gamma_{i} \gamma_{j} + \gamma_{j} \gamma_{i}  = -\frac12 g_{ij}
\eng
$g_{ij}( x)$ is the metric on the 3-sphere of radius $1/\epsilon$.
The $\SU(2)$ covariant derivative is
\eq
\label{eq:CGFcovariantderivative}
D_{i} = \nabla_{i}  +  b(z) \gamma_{i}(x)
\qquad
\nabla_{i} = \partial_{i}+ \epsilonb \gamma_{i}
\en
The action becomes
\eq
\label{eq:gaugeaction}
\frac1{\hbar} S_{\gauge} = 
\frac{6\pi^{2}  }{g^{2}\epsilonb^{3}}
\int
\bigg[-\frac12 \bigg(\frac{d b}{d z}\bigg)^{2}+ \frac12 
\left(b^{2}-\epsilonb^{2}\right)^{2}\bigg]
dz
\en
Local physics in the scaled coordinate system
is independent of $\epsilonb$
when $\epsilonb$ is very small.
So the value of the initial energy $\EhatCGF$ 
does not matter as long as it is very large.

The Jacobi
elliptic function $\cn(z,k)$ is doubly periodic in $z$.
The real period is $z \sim z +4\pi K$.
The imaginary period is $z \sim z +4\pi K'i$.
The numbers $K$ and $K'$ are the complete elliptic integrals of the first kind.
The CGF $\hat b(\that)$ oscillates with period
$t \sim t + 4Ka$ in real co-moving time.
The CGF is periodic in imaginary co-moving time with period
$t \sim t + 4K'a \,i$.
The period in imaginary time defines a temperature
$\kB T_{\CGF} = \hbar/( 4K'a)$.
The initial state of the \SM\ fluctuations
is specified by
requiring correlation functions to be periodic in imaginary time
with the period $ 4K'a i$.
The CGF acts 
as thermal bath for the fluctuations.
The initial state is specified by the five conditions 
above and the condition that the fluctuations are in the 
natural thermal state defined by the  CGF.

\section{The electro-weak transition}

The Higgs action is
\eqg
\label{eq:Higgsaction}
\frac1\hbar S_{\Higgs} =\int
\bigg [
 a^{-2} D_{\mu}\phi^{\dagger} D^{\mu}\phi
+  \frac12\lambdaH^{2}\bigg(\phi^{\dagger}\phi 
-  \frac{\vH ^{2}}2 \bigg)^{2}
\,\bigg]\, a^{4}\sqrt{-g} \,d^{4}x
\\[2ex]
 D_{\mu}\phi^{\dagger} D^{\mu}\phi
= \partial_{\mu}\phi^{\dagger} \partial^{\mu}\phi
+ \frac12 b(z) (\partial^{i}\phi^{\dagger} \gamma_{i}\phi
- \phi^{\dagger}\gamma_{i} \partial^{i}\phi)
+ \frac3{4} b(z)^{2} \phi^{\dagger}\phi
\eng
$\lambda$ is the Higgs coupling constant,
$\lambda^{2}=0.258$, and
$m_{\Higgs}=\hbar \lambda v = 125\,\GeV$ is the Higgs mass \cite{PDG2021}.
We have replaced  $\nabla_{i}=\partial_{i}+\epsilonb \gamma_{i}$
with $\partial_{i}$ assuming $\epsilonb$ to be very small.

The energy in the CGF and the 
Higgs field at $\phi=0$
drives an expanding universe.
The CGF oscillates much faster than
the expansion
so the adiabatic approximation
is accurate.
Averaging $b$ and $b^{2}$  over the
oscillation
gives the effective potential for $\phi$.
\eqg
\label{eq:effectivepotential}
V_{\eff}(\phi) 
= \frac{\lambdaH^{2}\vH^{4}}8
+ \left(
\frac3{4}  \frac{\expval{b^{2}} }{a^{2}}
-\frac{\lambdaH^{2}\vH ^{2}}2 
\right)
\phi^{\dagger}\phi
+ \frac{\lambdaH^{2}}2  (\phi^{\dagger}\phi)^{2}
\\[2ex]
\expval{b} = 0
\qquad
\expval{b^{2}} =
\frac1{4K}\int_{0}^{4K} k^{2} \cn^{2}(z,k) \; dz = \frac{\pi}{4 K^{2}}
\eng
At early times when $a(z)$ is small
the coefficient of $\phi^{\dagger}\phi$ is positive
so  $\phi=0$ is stable.
The coefficient of $\phi^{\dagger}\phi$ turns negative
when $a(z)$ reaches $a_{\EW}$,
\eqg
a_{\EW}
= \left(\frac{3 }{2} \frac{\expval{b^{2}}}{\lambdaH^{2}\vH^{2} }\right)^{\frac12}
=  \frac{(6\pi)^{\frac12}}{4 K \lambda v} 
= 0.5854 \, \frac{\hbar}{m_{\Higgs}}
= 3.08\ntimes 10^{-27}\,\sunit
\eng

\section{Classical solution after $a_{\EW}$}

After $a(z)=a_{\EW}$ the Higgs field moves away from $\phi=0$
towards its vacuum expectation value,
tracking the minimum of 
the effective potential (\ref{eq:effectivepotential}),
\eq
\label{eq:phisq}
(\phi^{\dagger}\phi)_{0}=
\frac{\vH ^{2}}2 
-
\frac3{4\lambdaH^{2}}  \frac{\expval{b^{2}} }{a^{2}}
\en
$\phi\neq 0$ breaks the $\Spin(4)$ symmetry but
the gauge field action remains $\Spin(4)$-symmetric
because,
for $B_{i}$ the $\SU(2)$ gauge field,
$\phi^{\dagger} B^{i}B_{i} \phi = \phi^{\dagger}  \phi
\, \tr (B^{i}B_{i})/2$.
The $\Spin(4)$-symmetric CGF continues to oscillate
but now with action and 
dimensionless energy 
\eqa
\label{eq:musq}
\frac1{\hbar} S^{\eff}_{\gauge} &= 
\frac{6\pi^{2} }{g^{2}\epsilonb^{3}}
\int
\bigg[-\frac12 \bigg(\frac{d b}{d z}\bigg)^{2}
+ \frac1{2} \mu^{2}b^{2} 
+ \frac12 b^{4}
\bigg]
dz
\\[2ex]
\ECGF &=  \frac12 \bigg(\frac{d b}{d z}\bigg)^{2}
+\frac{1}{2}\mu^{2}  b^{2}
+\frac12 b^{4}
\qquad
\mu^{2} = \frac12 \gtwo ^{2}a^{2}(\phi^{\dagger}\phi)_{0}
\ena
The adiabatic approximation remains valid.  
The oscillation changes slowly with 
$\mu^{2}$.
The classical solution is again a Jacobi elliptic function,
\eq
b(z) = \frac{ k \cn(u,k) }\alpha\qquad  dz = \alpha du
\qquad
\alpha^{2}\mu^{2}=1-2k^{2}
\qquad
\alpha^{4}\ECGF  =\frac{ k^{2}(1-k^{2})}{2}
\en
now parametrized by 
$k^{2}$ and $\alpha$
instead of $\mu^{2}$ and $E_{\CGF}$.
The solution $b(z)$ determines
the value of $\phi^{\dagger}\phi$ by  (\ref{eq:phisq}) and the 
identity
\eq
\alpha^{2}\expval{b^{2}} = \frac1{4K} \int_{0}^{4K}\alpha^{2} b^{2} \,du
= \frac1{4K} \int_{0}^{4K}  k^{2} \cn^{2}(u,k) \; du
= k^{2}-1 + \frac{E}{K} 
\en
$E$ is the complete elliptic integral of the second kind.
Equations (\ref{eq:phisq}) and (\ref{eq:musq}) combine to parametrize 
the scale $a$ by $k^{2}$.
\eq
\hat a = \frac{m_{\Higgs}a}{\hbar}
\qquad
\alpha^{2} \hat a^{2}
=
\frac32 \alpha^{2}\expval{b^{2}}
+
\frac{4\lambda^{2}}{\gtwo ^{2}}
\alpha^{2}\mu^{2}
\en
Finally, 
the adiabatic invariant
$\oint p\,dq$
is a constant of the motion
for an adiabatically evolving oscillator $q$.
The adiabatic  equation for the CGF is
\eq
\frac{(1-k^{2})K + (2k^{2}-1)E}{\alpha^{3}} = \text{ constant,
\quad with $\alpha=1$ at $k^{2}=1/2$.}
\en
The classical time evolution is now completely parametrized by $k^{2}$.
As $k^{2}$ evolves from $1/2$ to $0$,
the scale $a$ goes from 
$a_{\EW}$ to $\infty$ and  $\phi^{\dagger}\phi$ goes from $0$ to 
$v^{2}/2$.

\section{CGF equation of state}

The energy-momentum tensors of both the gauge field and $\phi$
are $\SO(4)$-symmetric so the CGF is a perfect fluid.
The total density $\rho_{\CGF} $ and total pressure $p_{\CGF}$ 
are obtained by substituting the classical 
solutions in the energy-momentum tensor.
Define the dimensionless density and pressure
\eqg
\hat \rho_{\CGF}(\hat a)  = \frac{\rho_{\CGF}}{\rho_{b}}
\qquad
\hat p_{\CGF}(\hat a)  = \frac{p_{\CGF}}{\rho_{b}}
\qquad
\rho_{b} = \frac{m_{\Higgs}^{4}}{\hbar^{3}}
= 5.68 \ntimes 10^{28} \, \frac{\kg}{\munit^{3}}
\eng
$\hat a = m_{\Higgs}a /\hbar$ is the dimensionless scale.
Before $a_{\EW}$
the gauge field is pure radiation.
and $\phi$ contributes vacuum energy.
\eq
\hat \rho_{\CGF}  = \frac{3 }{8 \gtwo ^{2}}\,\frac{1}{\hat a^{4}} 
+\frac{ 1}{8\lambdaH^{2}}
\qquad
\hat p_{\CGF}
= \frac{1 }{8 \gtwo ^{2}}\,\frac{1}{\hat a^{4}} 
-\frac{ 1}{8\lambdaH^{2}}
\en
After $a_{\EW}$,
\eq
\label{eq:equationofstatelow}
\hat\rho_{\CGF}(k^{2})
=  \frac1{\hat a^{4}} 
\left(
\frac{3\ECGF}{\gtwo ^{2}} + \frac{9\expval{b^{2}}^{2}}{32\lambda^{2}}
\right )
\quad
\hat p_{\CGF} (k^{2})
=  \frac1{\hat a^{4}}  \left(\frac{E_{\CGF}-\mu^{2} \expval{b^{2}} }{\gtwo ^{2}} 
- \frac{9\expval{b^{2}}^{2} }{32\lambdaH^{2} } \right)
\en
In both regimes the equation of state relating $p$ to $\rho$
is given implicitly by the two functions $\hat \rho$, $\hat p$.
The boundary between the two regimes is 
at  $\hat a_{\EW} =   0.585$, $\hat \rho_{\EW} = 7.97$.

\begin{figure}
\begin{center}
\captionsetup{justification=centering,margin=0.13\linewidth}
\includegraphics[scale=0.6]{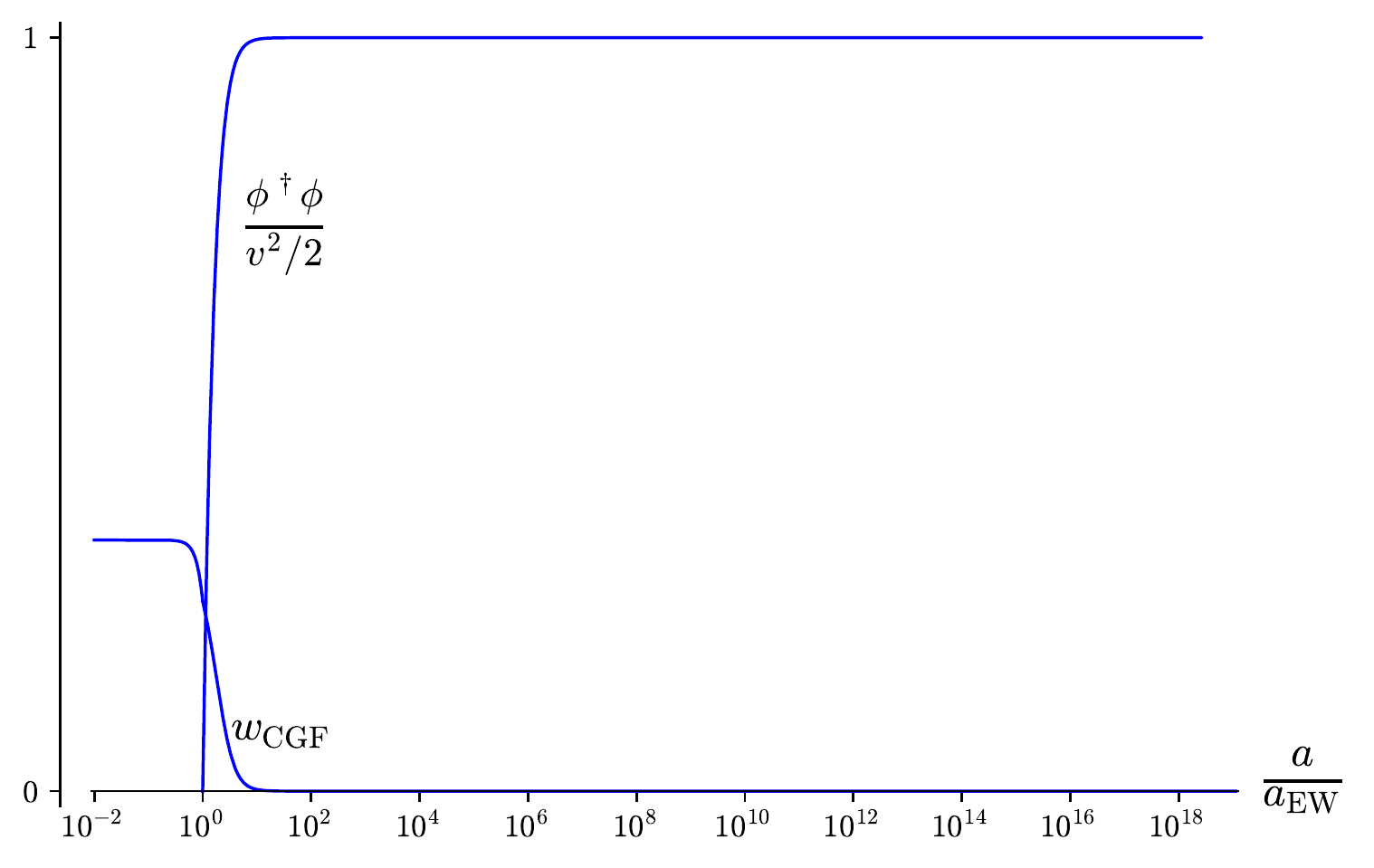}
\caption{
\label{fig:SMepoch}
}
\end{center}
\end{figure}

The equation of state parameter $w =  p/\rho$ is 
plotted in Figure~\ref{fig:SMepoch}.
The CGF behaves as a nonrelativistic perfect fluid, $w_{\CGF}\approx 
0$, from about $10\,a_{\EW}$ or $10^{2}\,a_{\EW}$ onward,
which is to say that the CGF is cold dark matter.

\section{Present flatness}

The Friedmann equation normalized by the Hubble constant is
\eqg
\label{eq:Friedmanneqnnormalized}
\frac{H^{2}}{H_{0}^{2}}  = \Omega_{m} + 
\Omega_{\ssLambda}+\Omega_{\curvature} 
\qquad
H= \frac1{a} \frac{da}{dt}= \frac1{a^{2}} \frac{da}{dz}
\\[1ex]
\rho_{\ssc} = \frac{3  H_{0}^{2}}{\kappa}
\qquad
\Omega_{m} = \frac{\rho_{m}}{\rho_{\ssc}}
\qquad
\Omega_{\ssLambda} = \frac{\rho_{\ssLambda}}{\rho_{\ssc}}
\qquad
-\Omega_{\curvature} 
=\frac{1}{H_{0}^{2}}\frac{1}{R^{2}}=\frac{ \epsilon^{2}}{H_{0}^{2} a^{2}}
\eng
The dark energy density is $\Omega_{\ssLambda}=0.685$
\cite{PDG2021}.
It is assumed due to the cosmological constant
so is constant in time.
The  present curvature is small,
$|\Omega_{\curvature}|<0.001$ \cite{PDG2021}.
The only matter
in the classical CGF universe 
is the CGF.
\eq
\label{eq:Omegam}
\rho_{\ssm} = \rho_{\CGF}
\qquad
\Omega_{m} = \Omega_{\CGF} = \frac{\rho_{\CGF}}{\rho_{\ssc}}
\en
The present time is identified by the condition $H=H_{0}$
which is $\Omega_{m}=0.315$.  
Solving (\ref{eq:Omegam}) 
gives  the present value
$k_{0}^{2}  = 7.89\ntimes 10^{-56}$ and
\eq
a_{0} =  4.54 \ntimes 10^{18} a_{\EW}
\qquad
-\Omega_{\curvature} 
=\frac{ \epsilon^{2}}{H_{0}^{2} a_{0}^{2}}
= 1.07\ntimes 10^{51} \; \epsilon^{2} 
\en
$|\Omega_{\curvature}| < 0.001$ is
$\epsilon  <  10^{-27}$
which is the initial energy condition
$\EhatCGF >  10^{107}$.

The dimensionless energy $\EhatCGF$ is the only adjustable 
parameter in the initial state.
If 
$\EhatCGF >  10^{107}$ then $\epsilon$ is so small that no local 
physics depends on the value of $\EhatCGF$.
In effect there are no adjustable parameters.

\section{Dark matter stars}

The classical CGF cosmology has no ordinary matter, only dark matter. 
The dark matter is the CGF.
The actual universe is a perturbation of this dark matter universe
by the fluctuations of the \SM\ fields.
Constructing the initial fluctuations and calculating their
time evolution is a well-defined calculation
within the \SM\ and \GR.
Detailed quantitative checks of the theory depend on that calculation,
which remains to be done.

\begin{figure}
\begin{center}
\includegraphics[scale=0.45]{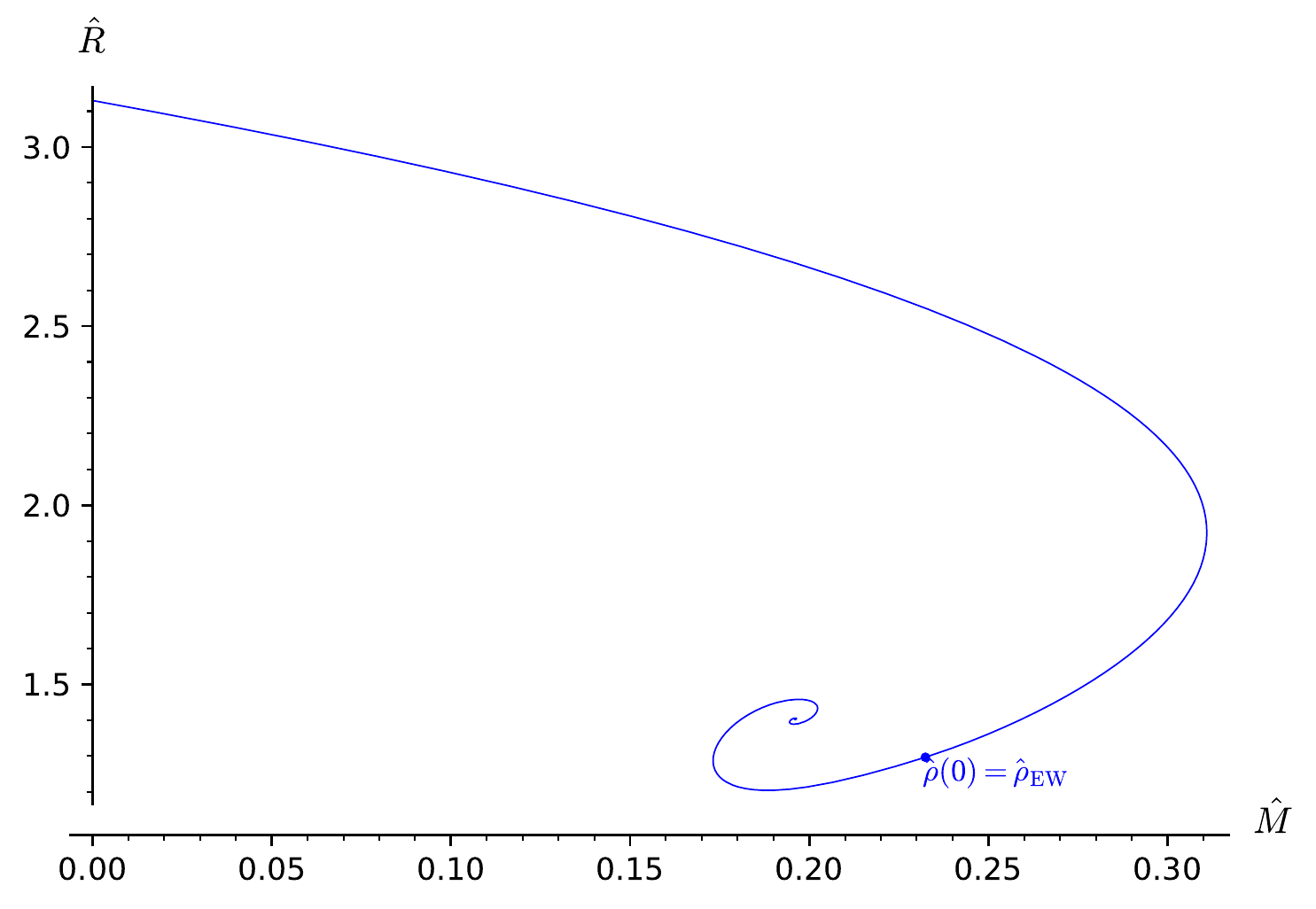}
\hfill
\includegraphics[scale=0.45]{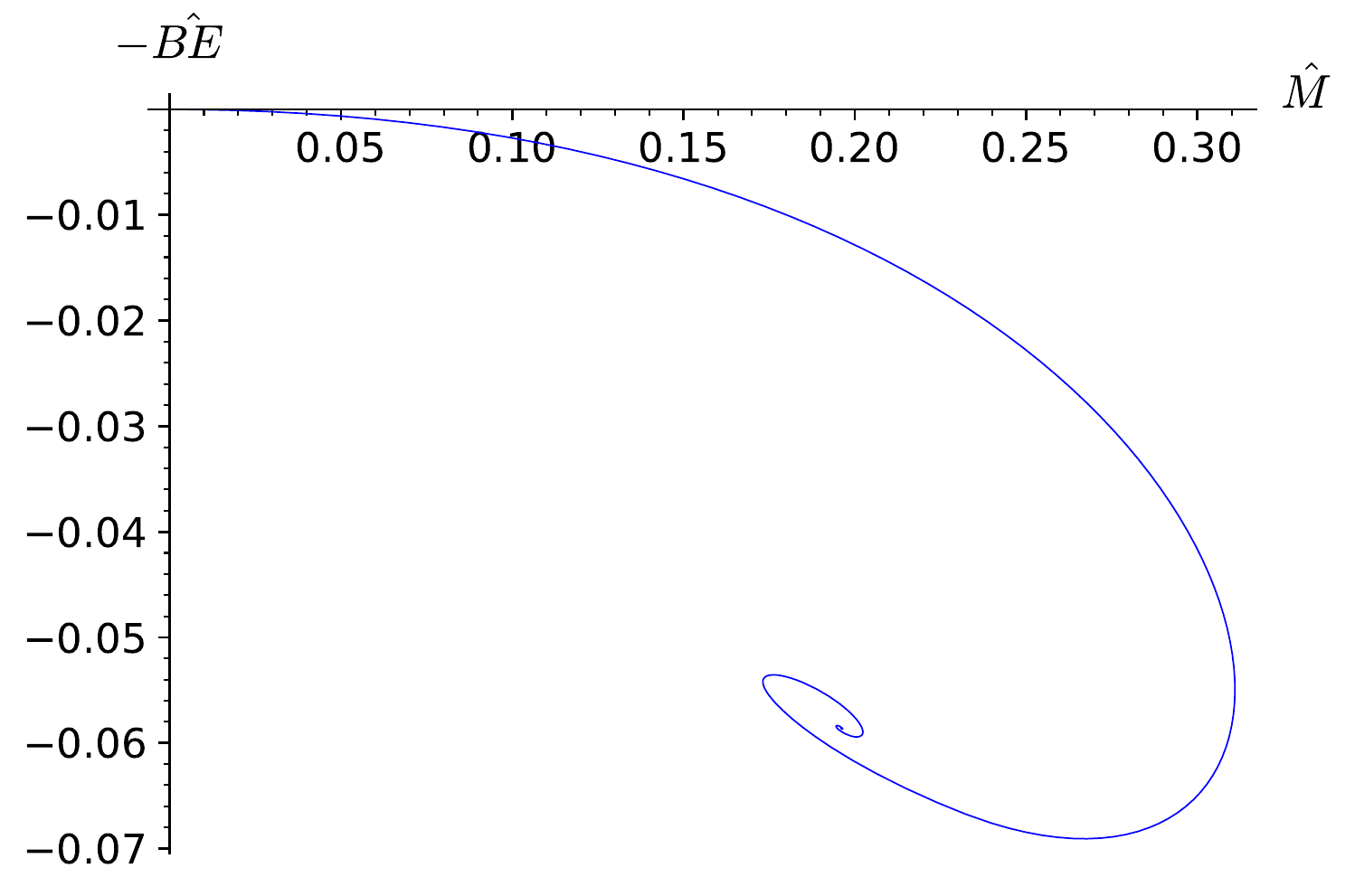}
\caption{\label{fig:MR}}
\end{center}
\end{figure}

Meanwhile, having in hand the equation of state of the classical CGF allows 
solving the TOV 
stellar structure equations to find the possible
stars made of the CGF.
Density fluctuations in the CGF presumably  
collapsed gravitationally to form self-gravitating 
bodies of which the simplest are the
spherically symmetric non-rotating stars governed by the TOV 
equations.
Figure~\ref{fig:MR} shows the results of solving the 
TOV equations numerically.
The gravitational scales are set by the CGF density scale $\rho_{b}=m_{\Higgs}^{4}/\hbar^{3}$.
\eqg
\rb  = (4\pi G \rhob)^{-1/2}= 4.34 \,\cmunit
\qquad
\mb = G^{-1} \rb = 2.94\ntimes 10^{-5} \Msun
= 5.26 \ntimes 10^{42}\,\text{J}
\\[1ex]
\hat R = \frac{R}{\rb}
\qquad
\hat M =\frac{M}{\mb}
\qquad
\hat{\mathrm{BE}} =\frac{\mathrm{BE} }{\mb}
\eng
The left plot is the mass-radius curve. 
The right plot is the 
mass-binding energy curve.
The curves spiral
inwards, parametrized by increasing central density.  

The dark matter universe is presumably populated with such stars.
The abundance distribution of their masses is a fluctuation calculation
still to be done.
Microlensing puts an upper limit 
$10^{-11}\Msun$
on such compact dark matter objects 
as the halos \cite{Niikura:2017zjd}.
So the halos must consist mostly of dark matter stars
of radius $R=13.6\,\cmunit$
at the low mass end of the curve.
It seems challenging to detect dark matter in such a form.
The rapidly oscillating CGF will have no 
significant non-gravitational interactions.

The binding energy curve shows the possibility of
metastable dark matter stars that could undergo explosive collapse
to smaller radius, emitting a burst of gravitation energy on the order of
$10^{41}$J in $10^{{-10}}$s.
Such bursts might be observable, perhaps taking place in the centers of ordinary stars
or out in the open.

Finally, a dark matter star near the asymptotic fixed point of the 
spiral has high central density.
Such objects could probe physics beyond the \SM.

\vskip2ex
I thank C.~Keeton 
for advice on microlensing and for suggesting reference \cite{Niikura:2017zjd}. 
This work was supported by the Rutgers New High Energy Theory Center
and by the generosity of B. Weeks.
I am grateful
to the  Mathematics Division of the 
Science Institute of the University of Iceland
for its hospitality.

\vspace*{-1.5ex}

\bibliographystyle{ytphys}
\raggedright
\phantomsection
\bibliography{Literature}

\end{document}